\documentclass[a4paper,12pt,oneside,tbtags]{article}

\usepackage{amsmath}            
\usepackage{amssymb}            
\usepackage{setspace}           
\usepackage{verbatim}           
\usepackage{graphicx}           
\usepackage{psfrag}             
\usepackage[nosort]{cite}

\makeatletter
\let\old@startsection=\@startsection
\renewcommand{\@startsection}[6]{\old@startsection{#1}{#2}{#3}{#4}{#5}{#6\mathversion{bold}}}
\makeatother

\newcommand{\bibtitle}[1]{{\em #1}}
\newcommand{\xxx}[1]{{\tt hep-th/#1}} 

\DeclareMathOperator{\tr}{tr}                               
\newcommand{\be}{\begin{equation}}
\newcommand{\ee}{\end{equation}}
\newcommand{\ba}{\begin{eqnarray}}
\newcommand{\ea}{\end{eqnarray}}
\newcommand{\gym}{g_{\scriptscriptstyle\mathrm{YM}}}        
\newcommand{\Ncal}{\mathcal{N}}                             
\newcommand{\R}{\mathbb{R}}                                
\newcommand{\unit}{\mathbf{1}}                              
\newcommand{\modulus}[1]{{| #1 |}}                          
\newcommand{\eps}{\varepsilon}                              
\newcommand{\comm}[2]{[#1,#2]}                              
\newcommand{\acomm}[2]{\{#1,#2\}}                           
\newcommand{\grSU}{\mathrm{SU}}                             
\newcommand{\grSO}{\mathrm{SO}}                             
\newcommand{\ket}[1]{\bigl|#1\bigr>}                        
\newcommand{\bra}[1]{\bigl<#1\bigr|}                        
\newcommand{\braket}[2]{\bigl<#1|#2\bigr>}                  

\newcommand{\Z}{{Z}}                                        
\newcommand{\Zb}{{\bar Z}}                                  
\newcommand{\W}{{\phi}}                                     
\newcommand{\Wb}{{\bar \phi}}                               

\begin{document}
\thispagestyle{empty}
\begin{flushright}
{\sc\footnotesize hep-th/0310232} \\
{\sc\footnotesize AEI-2003-086}
\end{flushright}
\vspace{1cm}
\setcounter{footnote}{0}
\begin{center}
{\Large{\bf On the Integrability of large \mathversion{bold}$N$ \mathversion{normal}\\
Plane-Wave Matrix Theory \par}}\vspace{20mm}
{\sc Thomas Klose and Jan Plefka} \\[7mm]
{\it Max-Planck-Institut f\"ur Gravitationsphysik \\
     Albert-Einstein-Institut \\
     Am M\"uhlenberg 1, D-14476 Potsdam, Germany} \\ [2mm]
{\tt thklose,plefka@aei.mpg.de} \\[10mm]

{\sc Abstract}\\[2mm]
\end{center}

\noindent We show the three-loop integrability of large $N$ plane-wave
matrix theory in a subsector of states comprised of two
complex light scalar fields. This is done by diagonalizing the
theory's Hamiltonian in perturbation theory and
taking the large $N$ limit. At one-loop level the result is known to
be equal to the Heisenberg spin-1/2 chain, which is a well-known
integrable system. Here, integrability implies the existence of 
hidden conserved charges and results in a degeneracy
of parity pairs in the spectrum. In order to confirm integrability at
higher loops, we show that this degeneracy is not lifted and that
(corrected) conserved charges exist. Plane-wave matrix theory is intricately
connected to $\Ncal=4$ Super Yang-Mills, as it arises as a consistent
reduction of the gauge theory on a three-sphere.
We find that after appropriately renormalizing the mass parameter of the plane-wave
matrix theory the effective Hamiltonian is identical to the dilatation
operator of $\Ncal=4$ Super Yang-Mills theory in the considered
subsector. Our results therefore represent a strong support for the
conjectured three-loop integrability of planar $\Ncal=4$ SYM and are in disagreement
with a recent dual string theory finding. Finally, we study the stability
of the large $N$ integrability against nonsupersymmetric
deformations of the model.

\newpage

\setcounter{page}{1}


\section{Introduction and Conclusions}

Recently a number of fascinating developments have pointed to
the existence of integrable structures in the large $N$ limit
of four-dimensional $\Ncal=4$ supersymmetric $\grSU(N)$ Yang-Mills 
theory \cite{Minahan,BKS,bes1,SYMint}. These developments have been paralleled by 
investigations on the integrability of the dual $AdS_5\times S^5$ 
string \cite{STRINGint}. In this paper we shall turn to another
setting of string/M-theory in which integrable structures
emerge: The large $N$ limit of plane-wave matrix theory \cite{bmn}.
This $\grSU(N)$ matrix quantum mechanics represents a mass deformed 
version of the ``BFSS'' matrix theory \cite{bfss}, thereby
preserving all of its supersymmetries. In its large $N$ limit
it is believed to describe the microscopic degrees of freedom 
of M-theory in the maximally supersymmetric plane-wave background 
\cite{Kow}. Opposed to its flat space cousin the model enjoys a number of  
pleasant features: It possesses a discrete spectrum, the model may 
be treated perturbatively in the large mass limit \cite{PWMM1}, and
its spectrum contains an infinite set of exactly protected states 
due to its underlying $\grSU(2|4)$ symmetry structure \cite{PWMM2,PWMM3,PWMM4}. 
The present work, however, is mainly motivated by the direct
connection of plane-wave matrix theory to $\Ncal=4$ Super 
Yang-Mills on $\R\times S^3$. The matrix quantum
mechanics arises in a consistent reduction of the four-dimensional
superconformal gauge theory on the $S^3$ \cite{us}. As the scaling dimensions
of operators in a conformal theory on $\R^4$ are related to the
energy spectrum on $\R\times S^3$ in radial quantization, the
dilatation operator of the four-dimensional Super Yang-Mills
theory should bear a relation to the Hamiltonian of the plane-wave matrix model. 
And indeed in \cite{us} it was shown that at
the one-loop level the spectrum of anomalous dimensions of gauge theory
operators agrees with the leading perturbative corrections to the
spectrum of associated states in the plane-wave matrix quantum mechanics.
An immediate consequence of this observation is, that the one-loop 
integrability of the Super Yang-Mills model, to be discussed below, 
carries over to plane-wave matrix theory.

Integrability in $\Ncal=4$ Super Yang-Mills
shows up in the perturbative study of the dilatation operator,
whose eigenvalues yield the spectrum of scaling dimensions 
of composite operators. In principle the scaling dimensions
are extracted from diagonal two-point functions, however, in
\cite{Beisert:2002ff,Minahan} a new efficient technique was introduced to directly
extract the dilatation operator order by order in a loop
expansion from the gauge theory.  This led to the remarkable
discovery of Minahan and Zarembo \cite{Minahan} of an exact equivalence
between the planar, one-loop piece of the dilatation operator (in the 
pure scalar excitation sector) 
with the Hamiltonian of an integrable $SO(6)$ spin chain. In this
picture the spins on the chain are given by the scalar fields
$\phi_i$ ($i=1,\ldots,6$) of $\Ncal=4$ Super Yang-Mills 
thread together in a single trace  structure, 
$\tr(\phi_{i_1}\ldots \phi_{i_n})$. This finding was subsequently
generalized to the full Super Yang-Mills field content
\cite{bes1} establishing an equivalence of the planar one-loop
dilatation operator with the Hamiltonian of an integrable 
$\grSU(2,2|4)$ spin chain.

Obviously one would like to know whether
this integrability is a mere artifact of the one-loop 
approximation, or whether
it represents a genuine hidden symmetry of planar Super Yang-Mills.
To answer this question higher loop considerations suggest themselves.
For this it is advantageous to restrict one's attention to a minimal
subsector of field excitations, which does not mix in perturbation
theory with other excitations. 
Such a sector is given by the $\grSU(2)$ excitations of two
complex scalars, which we denote $Z$ and $\phi$.
The corresponding
one-loop planar dilatation operator is then identical to the famous
XXX${}_{1/2}$ Heisenberg quantum spin chain model, $H=\sum_{j=1}^L \vec s_j\cdot
\vec s_{j+1}$, where $\vec s_j$ is the spin $1/2$ operator acting
on the $j$th site of the chain, being in an up ($Z$) or down ($\phi$)
state. This model of a one dimensional
magnet is known to be integrable, i.e.~one
may construct a set of $L-1$  mutually commuting charges which
contain $H$ \cite{Faddeev:1996iy}. 
In \cite{BKS} a two-loop computation in this $\grSU(2)$
sector of $\Ncal=4 $ Super Yang-Mills
was performed showing that integrability is
preserved. The higher loop contributions to the planar dilatation operator
result in deformations of the spin chain Hamiltonian
incorporating spin interactions beyond nearest neighbours, i.e.~at
$l$ loop level one is dealing with the interactions of $l+1$ spins.
The integrability of the higher loop deformation manifests itself
in the preservation of degeneracies in the spectrum, which may be attributed
at one-loop level to the hidden symmetry. However, a generalization
of the powerful methods known from integrable spin chains for the construction of
commuting charges at higher loop orders
remains unknown, nevertheless the first few charges beyond the deformed
Hamiltonian were constructed explicitly. It is important to note that while the one-loop
dilatation operator is exactly integrable, the combined one- and 
two-loop dilatation operator is only integrable up to terms of
three-loop order. Hence only the full, planar, all-loop
dilatation operator can display exact integrability -- a fascinating 
perspective on the large $N$ spectrum of $\Ncal=4$ Super Yang-Mills.

Based on the assumption of {\it three}-loop integrability the 
authors of \cite{BKS} where actually able to determine the form of
the planar {\it three}-loop dilatation operator.
However, this conjecture was
recently challenged by the work of Callan, Lee, McLoughlin, Schwarz,
Swanson and Wu \cite{Callan:2003xr} from
the dual string theory side. These authors computed the corrections
to the spectrum of IIB plane-wave superstring theory due to
the presence of first curvature corrections that arise in the
expansion of $AdS_5\times S^5$ about the plane-wave limit. These
corrections translate into certain $1/J$ corrections of the scaling
dimensions of BMN operators in the BMN limit \cite{bmn} and can be compared to
the two-loop result and three-loop conjecture of Beisert, Kristjansen 
and Staudacher \cite{BKS}.
Indeed the two-loop corrections of the gauge theory match with
the string result, however, at three-loop order the conjectured
gauge theory result {\it disagrees} with the string based findings.
Hence, if the involved analysis of \cite{Callan:2003xr} is indeed correct, then
string theory predicts the breakdown of integrability in 
$\Ncal=4$ Super Yang-Mills at the three-loop level! 
To clarify this issue from a gauge theory perspective a direct three-loop
analysis would be very desirable. Fortunately the aforementioned relation
of $\Ncal=4$ Super Yang-Mills to plane-wave matrix theory enables
one to perform such a three-loop computation in a simpler quantum
mechanical setting. This is the goal of our paper.

Our analysis consists of an essentially straightforward application
of degenerate quantum mechanical perturbation theory to the third non-vanishing 
order, which is equivalent to a field theoretic three-loop computation.
Needless to say that this has been only possible by making massive use of
symbolic computer algebra systems.  
The outcome of our analysis is positive for integrability: Plane-wave 
matrix theory remains integrable up to the three-loop level 
in the minimal $\grSU(2)$ sector mentioned above. 
Technically this means that the degeneracy of ``planar parity pairs''
noted in \cite{BKS} extends to the three-loop order. Also the first
charge beyond the Hamiltonian can be explicitly constructed.
Clearly this does not
prove the integrability of the full four-dimensional theory, 
as we are dealing with a reduced model here, but
it does make it very likely. This is furthermore
supported by the following
remarkable observation. As found in \cite{us} the agreement of
the Super Yang-Mills dilatation operator with the perturbative
effective plane-wave matrix theory Hamiltonian ceases to exist at two-loop
order. However, if one allows for a renormalization of the relation between the
(dimensionless) mass parameter $M$ of matrix theory and the
Yang-Mills coupling constant $\gym$ of the form\footnote{This observation grew
out of discussions with N.~Beisert and M.~Staudacher.} 
\be
\frac{1}{M^3}=\frac{\gym^2}{32\pi^2}\, \left ( 1+\frac{7}{32\,\pi^2}\,
\gym^2\, N - \frac{11}{252\, \pi^4}\,\gym^4\, N^2+{\cal O}(\gym^6)\, \right )\, ,
\ee
then this agreement extends to three-loop order! That is our three-loop
matrix theory computation reproduces the conjectured three-loop dilatation
operator of Beisert, Kristjansen and Staudacher \cite{BKS} 
with the above choice of renormalized
mass parameter. Indeed such a renormalization is to be expected if
one were to truly path integrate out the higher Kaluza-Klein modes of the
gauge theory on $S^3$ 
to obtain an effective one dimensional theory
of the zero-modes.
The resulting effective theory would then be given by the plane-wave matrix model plus
higher loop corrections in $\lambda=\gym^2\, N$ and $1/N$. These quantum loop
corrections should manifest themselves  in a renormalization
of the mass parameter $M$ of the plane-wave matrix model 
and -- but our findings indicate that this is not the case
-- the addition of higher order interaction terms.
It would be very interesting to study this in more detail, in
particular the r\^ole of (maximal) supersymmetry in this non-emergence of
higher order interaction terms.

In addition to that, we
have explored how stable the integrability is to deformations
of the theory. It turns out that supersymmetry is absolutely {\it unessential}
for integrability in the matrix model up to three-loops. Switching off all fermions 
in the plane-wave matrix theory
does not destroy integrability. Furthermore
deformations of the bosonic field content are also possible without loosing
integrability, as long as the structure of Yang-Mills type interactions
is maintained. Our findings suggest that the hidden 
integrability, if it exists to all loops,  is a genuine property of large $N$ gauge
quantum mechanics. It would be very nice to prove integrability in these models.
To this end a formalism enabling one to directly work in the planar limit of 
the theory is needed. Maybe the structures introduced in \cite{halpern}
could be of use here.

Naturally a further important question is to understand how 
these (potential) integrable structures are to be interpreted in M-theory.
 
Finally, let us mention that
an alternative method to address the question of three-loop integrability 
of the full four-dimensional $\Ncal=4$ gauge theory
lies in the study of how strongly the underlying $\grSU(2,2|4)$ superalgebra 
-- or a suitable subgroup thereof -- restricts the three-loop structure 
of the dilatation operator. This will be addressed in \cite{Niklas}.


\section{Plane-wave Matrix Theory}

The Hamiltonian of plane-wave matrix theory can be written as $H = H_0 + V_1 + V_2$ with
\begin{subequations} \label{hamiltonian}
\begin{align}
  H_0 & = \tr \left[ \tfrac{1}{2} P_I P_I + \tfrac{1}{2} \left( \tfrac{M}{2} \right)^2 X_i X_i + \tfrac{1}{2} M^2 X_{i'} X_{i'} - \tfrac{3M}{4} i \theta \gamma_{123} 
\theta \right] \; , \\
  V_1 & = - M i\eps_{i'j'k'} \tr X_{i'} X_{j'} X_{k'} - \tr \theta \gamma_I \comm{X_I}{\theta} \; ,\\
  V_2 & = - \frac{1}{4} \tr \comm{X_I}{X_J} \comm{X_I}{X_J} \; .
\end{align}
\end{subequations}
The degrees of freedom are the light scalars $X_i$ ($i=1,\ldots,6$) of mass $M/2$, the heavy scalars $X_{i'}$ ($i'=1,\ldots,3$) of mass $M$ and a $\grSO(9)$ Majorana 
spinor $\theta_\alpha$ ($\alpha=1,\ldots,16$) of mass $3M/4$
\footnote{The mass parameter here is a dimensionless quantity given by $M=\frac{\mu\, l_P^2}{6\, R}$ in terms
of the M-theoretic quantities $\mu$ (mass parameter of the plane-wave background, see eq.~(1) of \cite{us}), $l_P$ (eleven dimensional Planck length)
and $R$ (radius of compact eleventh dimension).}.
All fields are $N\times N$ traceless Hermitian matrices 
by virtue of the gauge group $\grSU(N)$. The index $I=(i,i')$ embraces all scalars. In the fermionic sector we work in a representation with charge conjugation matrix 
equal to unity and with symmetric  euclidean Dirac matrices $(\gamma_{I})_{\alpha\beta}$. Note that it is the bosonic $X_i$ sector which is intimately
related to the scalar field sector of $\Ncal=4$ supersymmetric Yang-Mills theory.

We introduce the following creation and annihilation operators
\begin{align}
  P_i           & = \sqrt{\tfrac{M}{4}}  \left( a_i    + a_i^\dag    \right) &
  P_{i'}        & = \sqrt{\tfrac{M}{2}}  \left( b_{i'} + b_{i'}^\dag \right) &
                & \theta_\alpha = \theta^+_\alpha + \theta^-_\alpha \nonumber    \\
  X_i           & = \tfrac{i}{\sqrt{M}}  \left( a_i    - a_i^\dag    \right) &
  X_{i'}        & = \tfrac{i}{\sqrt{2M}} \left( b_{i'} - b_{i'}^\dag \right) &
                & \mbox{with } \theta^\pm := \Pi^\pm \theta \; , 
\end{align}
where $\Pi^\pm := \tfrac{1}{2}(\unit \pm i\gamma_{123})$. They satisfy the canonical (anti-)commutation relations
\begin{align}
  \comm{(a_i)_{rs}}{(a_j)_{tu}}       & = \delta_{ij}   \left( \delta_{st} \delta_{ru} - \tfrac{1}{N} \delta_{rs} \delta_{tu} \right) \; , \nonumber\\
  \comm{(b_{i'})_{rs}}{(b_{j'})_{tu}} & = \delta_{i'j'} \left( \delta_{st} \delta_{ru} - \tfrac{1}{N} \delta_{rs} \delta_{tu} \right) \; , \nonumber\\
  \acomm{(\theta^-_\alpha)_{rs}}{(\theta^+_\beta)_{tu}} & = \tfrac{1}{2} \Pi^-_{\alpha\beta} \left( \delta_{st} \delta_{ru} - \tfrac{1}{N} \delta_{rs} \delta_{tu} 
\right) 
\end{align}
and the free Hamiltonian now reads
\be
  H_0 = \tr \left[ \frac{M}{2} a_i^\dag a_i + M b_{i'}^\dag b_{i'} + \frac{3M}{2} \theta_\alpha^+ \theta^-_\alpha \right] \; .
\ee
Physical states are constraint to be gauge invariant and are given by traces over words in the
creation operators.


\section{Perturbation Theory}

We are faced with an eigenvalue problem of the form
\be \label{eqn:eigenvalue_problem}
  (H_0 + \kappa V_1 + \kappa^2 V_2) \ket{\phi(\kappa)} = E(\kappa) \ket{\phi(\kappa)},
\ee
where the spectrum of the free $(\kappa=0)$ Hamiltonian $H_0$ is known. Let us sketch how \eqref{eqn:eigenvalue_problem} is solved perturbatively in $\kappa$ and how 
we deal with potential degeneracies. Say we are interested in the perturbative 
shift of a specific $L$-fold degenerate free energy value $E_0$. Then we choose an arbitrary basis $\{\ket{\phi_i},\, i=1,\ldots,L\}$ for all states with this free 
energy
\be \label{eqn:free_eigenvalue_equation}
  H_0 \ket{\phi_i} = E_0 \ket{\phi_i} \qquad (i=1,\ldots,L)\; .
\ee
For later convenience we do not assume this basis to be orthonormal. 
All other states are labeled by $\ket{n}$ and their free energies by $E_n$. These energies may, of course, be degenerate as well. Here, however, we define the states 
in such a way that
\be
  \braket{n}{m} = \delta_{nm} \quad , \quad  \braket{n}{\phi_i} = 0 \; , 
\ee
which is always possible. 

When the interaction is turned on $(\kappa\not=0)$, the states $\ket{\phi_i}$ get corrected to $\ket{\phi_i(\kappa)}$. Clearly, the corrected states 
$\ket{\phi_i(\kappa)}$ are generically not eigenstates of the full Hamiltonian $H = H_0 + \kappa V_1 + \kappa^2 V_2$ since there may occur mixing among the states 
$\ket{\phi_i}$. However, we do not want to deal with mixing problems and the possible lift of the original degeneracy at this stage, and therefore introduce an energy 
matrix $E_{ij}(\kappa)$:
\be \label{eqn:degenerate_eigenvalue_equation}
  (H_0 + \kappa V_1 + \kappa^2 V_2) \ket{\phi_i(\kappa)} = E_{ij}(\kappa) \ket{\phi_j(\kappa)} \; .
\ee
Later, this matrix can be diagonalized and its eigenvalues are the final corrected energies. In this way we have decoupled the mixing of states in the degenerate 
subspace from the admixture of states from outside this subspace, which turns out to be extremely convenient for practical computations. 

Now we wish to determine an energy operator $T(\kappa)$ whose matrix elements give exactly the energy matrix:
\be \label{eqn:definition_energy_operator}
  E_{ij}(\kappa) = \bra{\tilde\phi_j} T(\kappa) \ket{\phi_i} \; .
\ee
In \eqref{eqn:definition_energy_operator} we have introduced the dual basis $\{\ket{\tilde\phi_i},\, i=1,\ldots,L\}$ defined by
\be
  \braket{\tilde\phi_i}{\phi_j} = \delta_{ij} \; .
\ee 
This is necessary because we did not demand the orthonormality of $\{\ket{\phi_i}\}$. On the other hand it is also very convenient for the following reason. We can 
start from any arbitrary basis $\{\ket{\phi_i}\}$ and do not need to worry about orthogonalizing it but may immediately apply $T(\kappa)$ with a result of the form
\be \label{eqn:energy_operator_applied}
  T(\kappa) \ket{\phi_i} = \sum_j t_{ij}(\kappa) \ket{\phi_j} + \sum_n t_{in}(\kappa) \ket{n} \; .
\ee 
From the expansion coefficients we then simply read off the energy matrix as
\be
  E_{ij}(\kappa) = t_{ij}(\kappa) \; .
\ee 

In order to find $T(\kappa)$ we follow a standard procedure. States and energy matrix are expanded in $\kappa$:
\begin{align}
  \ket{\phi_i(\kappa)} & = \ket{\phi_i} + \kappa   \left( \varphi^{(1)}_{ij} \ket{\phi_j} + \psi^{(1)}_{in} \ket{n} \right)
                                        + \kappa^2 \left( \varphi^{(2)}_{ij} \ket{\phi_j} + \psi^{(2)}_{in} \ket{n} \right)
                                        + \ldots \\
  E_{ij}(\kappa)       & = E^{(0)}_{ij} + \kappa E^{(1)}_{ij} + \kappa^2 E^{(2)}_{ij} + \ldots \qquad\mbox{with $E^{(0)}_{ij} = E_0 \delta_{ij}$}  
\label{eqn:expansion_energy_matrix}
\end{align}
Plugging these expansions into \eqref{eqn:degenerate_eigenvalue_equation} yields one equation for each power of $\kappa$. When the resulting equations are projected 
onto $\bra{\tilde\phi_i}$ and $\bra{n}$ one obtains expressions for $E_{ij}^{(k)}$ and $\psi_{in}^{(k)}$. The coefficients $\varphi_{ij}^{(k)}$ remain undetermined 
and may be set to zero. Iteratively we then find the energy operator
\be
  T(\kappa) = \sum_{k=0}^\infty \kappa^k T_k
\ee
up to third loop order 
\begin{subequations} \label{eqn:energy_operator}
\begin{align}
T_0  = &\ H_0 \\
T_2  = &\ V_1 \Delta V_1 + V_2 \\
T_4  = &\ V_1 \Delta V_1 \Delta V_1 \Delta V_1 + V_1 \Delta V_1 \Delta V_2 + V_1 \Delta V_2 \Delta V_1 + V_2 \Delta V_1 \Delta V_1 + V_2 \Delta V_2 \\
       & \quad - V_1 \Delta^2 V_1 P T_2 \nonumber
\end{align}
\be
\begin{split}
T_6 = &\  V_1 \Delta V_1 \Delta V_1 \Delta V_1 \Delta V_1 \Delta V_1 \\
      & + V_1 \Delta V_1 \Delta V_1 \Delta V_1 \Delta V_2
        + V_1 \Delta V_1 \Delta V_1 \Delta V_2 \Delta V_1
        + V_1 \Delta V_1 \Delta V_2 \Delta V_1 \Delta V_1 \\
      & + V_1 \Delta V_2 \Delta V_1 \Delta V_1 \Delta V_1
        + V_2 \Delta V_1 \Delta V_1 \Delta V_1 \Delta V_1 \\
      & + V_1 \Delta V_1 \Delta V_2 \Delta V_2
        + V_1 \Delta V_2 \Delta V_1 \Delta V_2
        + V_1 \Delta V_2 \Delta V_2 \Delta V_1 \\
      & + V_2 \Delta V_1 \Delta V_1 \Delta V_2
        + V_2 \Delta V_1 \Delta V_2 \Delta V_1
        + V_2 \Delta V_2 \Delta V_1 \Delta V_1 \\
      & + V_2 \Delta V_2 \Delta V_2 \\
      & - \bigl( V_1 \Delta^2 V_1 \Delta V_1 \Delta V_1
               + V_1 \Delta V_1 \Delta^2 V_1 \Delta V_1
               + V_1 \Delta V_1 \Delta V_1 \Delta^2 V_1 \\
      & \quad\;+ V_1 \Delta^2 V_1 \Delta V_2
               + V_1 \Delta^2 V_2 \Delta V_1
               + V_2 \Delta^2 V_1 \Delta V_1 \\
      & \quad\;+ V_1 \Delta V_1 \Delta^2 V_2
               + V_1 \Delta V_2 \Delta^2 V_1
               + V_2 \Delta V_1 \Delta^2 V_1 \\
      & \quad\;+ V_2 \Delta^2 V_2                 \bigr) P T_2 \\
      & - V_1 \Delta^2 V_1 P T_4
        + V_1 \Delta^3 V_1 P T_2 P T_2
\end{split}
\ee
\end{subequations}
giving rise to the expansion of the energy matrix \eqref{eqn:expansion_energy_matrix}.
Moreover $T_k$ vanishes for odd $k$. Here we have introduced the ``propagator'' and the projector
\be
  \Delta = \frac{1 - P}{E_0 - H_0} \qquad\mbox{and}\qquad P = \sum_{i=1}^L \ket{\phi_i}\bra{\tilde\phi_i} \; ,
\ee
respectively, where $P$ projects onto the subspace of states with free energy $E_0$.


\section{Energy Operator in $\grSU(2)$ subsector}

In this section we compute the energy operator $T$ up to third loop order for the $\grSU(2)$ subsector of the matrix model which is spanned by gauge invariant states 
composed of the two complex fields
\begin{align}
  \Z = \frac{1}{\sqrt{2}}(a_1^\dag + i a_2^\dag)  \qquad\mbox{and}\qquad  \W = \frac{1}{\sqrt{2}}(a_3^\dag + i a_4^\dag) \; .
\end{align}
The arguments given in section 3.1.~of\cite{BKS} for the group theoretical constraints on mixing 
of the analogue $\grSU(2)$ sector in $\Ncal=4$ supersymmetric Yang-Mills field theory are
inherited by the plane-wave matrix theory. This is so as these arguments are merely based on the representation
content of the excitations with respect to the $\grSO(6)$ $R$-symmetry group and the classical scaling dimensions 
of the involved fields, which directly carry over to the matrix model. In terms of the $\grSO(6)$ Dynkin
labels the states we consider transform in the representations $[p,q,p]$ and have free energy $E_0=\frac M 2\,(2p+q)$.

We apply a procedure similar to our previous two-loop computation \cite{us}. We normal order the terms given in \eqref{eqn:energy_operator} and discard all pieces 
which annihilate multi-trace states of $\Z$ and $\W$. Furthermore we can neglect all terms which are not relevant for the expansion coefficients $t_{ij}$ but only for 
$t_{in}$ in \eqref{eqn:energy_operator_applied}. The relevant terms are exactly those which conserve the free energy, i.e. the number of fields of a state.

In addition to the above fields, the energy operator consists also of the canonically conjugated fields
\begin{align}
  \Zb = \frac{1}{\sqrt{2}}(a_1 - i a_2)  \qquad\mbox{and}\qquad  \Wb = \frac{1}{\sqrt{2}}(a_3 - i a_4) \; ,
\end{align}
which satisfy
\be
  \comm{\Zb_{rs}}{\Z_{tu}} = \comm{\Wb_{rs}}{\W_{tu}} = \delta_{st} \delta_{ru} - \tfrac{1}{N} \delta_{rs} \delta_{tu} \; .
\ee
Since our intention is to compare the energy operator of the matrix theory to the dilatation generator of the field theory, we make the following definitions. 
Perturbation theory is done for large masses in an expansion in $1/M^3$. As shown in \cite{us}, this mass parameter should be related to the SYM coupling constant 
$\gym$ by
\be \label{eqn:relation_coupling_constants_oneloop}
  \frac{1}{M^3} = \frac{\gym^2}{32\pi^2} \;\; \left( = \frac{1}{2} G^2 = \frac{\Lambda}{2N} \right) 
\ee
in order to match all one-loop results in the scalar $\grSO(6)$ sector of both theories. In \eqref{eqn:relation_coupling_constants_oneloop} we have also given the 
relation to two further shorthands $\displaystyle G := \frac{\gym}{4\pi}$ and $\Lambda := G^2 N$. Now we define the counterpart of the dilatation generator from the 
energy operator $T$ by factoring out the free energy $M/2$ of an $\grSO(6)$ scalar, i.e. the mass of $\Z$ and $\W$:
\be
  T(\Lambda) =: \frac{M}{2} D(\Lambda) =: \frac{M}{2} \sum_{k=0}^\infty \Lambda^k D_{2k} \; ,
\ee
which explicitly  means
\begin{align}
  D_0 & = \frac{2}{M}      T_0 \; , &
  D_2 & = \frac{M^2}{N}    T_2 \; , &
  D_4 & = \frac{M^5}{2N^2} T_4 \; , &
  D_6 & = \frac{M^8}{4N^3} T_6 \; .
\end{align}
We will refer to $D(\Lambda)$ as the dilatation operator and to its matrix elements $\bra{\tilde\phi_j} D(\Lambda) \ket{\phi_i}$ as the dilatation matrix. The 
eigenvalues of $D(\Lambda)$ will be denoted as
$\Delta(\Lambda)$.

Performing the computation using {\tt Form} \cite{Jos} as well as {\tt Mathematica}, we find the following result:
\begin{subequations} \label{eqn:dilatation_operator}
\begin{align}
  D_0 =\ &                 : \tr (\Z\Zb + \W\Wb) : \; , \\
  D_2 =\ &    -\frac{2}{N} : \tr \comm{\Z}{\W} \comm{\Zb}{\Wb} : \; ,\\
  D_4 =\ &    \frac{11}{N} : \tr \comm{\Z}{\W} \comm{\Zb}{\Wb} : \nonumber \\
         & + \frac{2}{N^2} : \bigl( \tr \comm{\Z}{\W} \comm{\Zb}{\comm{\Z}{\comm{\Zb}{\Wb}}}
                                  + \tr \comm{\Z}{\W} \comm{\Wb}{\comm{\W}{\comm{\Zb}{\Wb}}} \bigr) : \; ,
\end{align}
\be
\begin{split}
  D_6 =  - \frac{115}{N}                          : &\tr \comm{\Z}{\W} \comm{\Zb}{\Wb} : \\
         + \frac{61}{8N^3}                 : \bigl( &\tr \comm{\comm{\comm{\W}{\Z}}{T^a}}{\Z} \comm{\Zb}{\comm{T^a}{\comm{\Wb}{\Zb}}}
                                                   + \tr \comm{\comm{\comm{\W}{\Z}}{T^a}}{\W} \comm{\Wb}{\comm{T^a}{\comm{\Wb}{\Zb}}} \bigr) : \\
         - \frac{9}{16N^3} \Bigl[          : \bigl( &\tr \Z \comm{T^a}{\comm{\Zb}{\comm{\Zb}{\comm{T^a}{\comm{\Wb}{\comm{\W}{\Z}}}}}}
                                                   + \tr \W \comm{T^a}{\comm{\Wb}{\comm{\Wb}{\comm{T^a}{\comm{\Zb}{\comm{\Z}{\W}}}}}} \bigr) : \\
                         - \tfrac{111}{31} : \bigl( &\tr \Z \comm{T^a}{\comm{\Zb}{\comm{\Wb}{\comm{T^a}{\comm{\Zb}{\comm{\W}{\Z}}}}}}
                                                   + \tr \W \comm{T^a}{\comm{\Wb}{\comm{\Zb}{\comm{T^a}{\comm{\Wb}{\comm{\Z}{\W}}}}}} \bigr) : \\
                         + \tfrac{80}{31}  : \bigl( &\tr \Z \comm{T^a}{\comm{\Wb}{\comm{\Zb}{\comm{T^a}{\comm{\Zb}{\comm{\W}{\Z}}}}}}
                                                   + \tr \W \comm{T^a}{\comm{\Zb}{\comm{\Wb}{\comm{T^a}{\comm{\Wb}{\comm{\Z}{\W}}}}}} \bigr) : \Bigr] \\
         + \frac{195}{16N^3} \Bigl[        : \bigl( &\tr \Zb \comm{T^a}{\comm{\Z}{\comm{\Z}{\comm{T^a}{\comm{\W}{\comm{\Wb}{\Zb}}}}}}
                                                   + \tr \Wb \comm{T^a}{\comm{\W}{\comm{\W}{\comm{T^a}{\comm{\Z}{\comm{\Zb}{\Wb}}}}}} \bigr) : \\
                         - \tfrac{111}{31} : \bigl( &\tr \Zb \comm{T^a}{\comm{\Z}{\comm{\W}{\comm{T^a}{\comm{\Z}{\comm{\Wb}{\Zb}}}}}}
                                                   + \tr \Wb \comm{T^a}{\comm{\W}{\comm{\Z}{\comm{T^a}{\comm{\W}{\comm{\Zb}{\Wb}}}}}} \bigr) : \\
                         + \tfrac{80}{31}  : \bigl( &\tr \Zb \comm{T^a}{\comm{\W}{\comm{\Z}{\comm{T^a}{\comm{\Z}{\comm{\Wb}{\Zb}}}}}}
                                                   + \tr \Wb \comm{T^a}{\comm{\Z}{\comm{\W}{\comm{T^a}{\comm{\W}{\comm{\Zb}{\Wb}}}}}} \bigr) : \Bigr] \\
         - \frac{2}{N^3}                          : &\tr \comm{\W}{\Z} \comm{\comm{\comm{\Zb}{\Wb}}{\W}}{\comm{\Z}{\comm{\Zb}{\Wb}}} : \\
         + \frac{4}{N^3}                   : \bigl( &\tr \comm{\comm{\comm{\Wb}{\Zb}}{\Z}}{\Zb} \comm{\Z}{\comm{\Zb}{\comm{\Z}{\W}}}
                                                  +  \tr \comm{\comm{\comm{\Wb}{\Zb}}{\W}}{\Wb} \comm{\W}{\comm{\Wb}{\comm{\Z}{\W}}} \\
                                                  + &\tr \comm{\comm{\comm{\Wb}{\Zb}}{\W}}{\Wb} \comm{\Z}{\comm{\Zb}{\comm{\Z}{\W}}}
                                                  +  \tr \comm{\comm{\comm{\Wb}{\Zb}}{\Z}}{\Zb} \comm{\W}{\comm{\Wb}{\comm{\Z}{\W}}} \bigr) : \\
         - \frac{2}{N^3}                   : \bigl( &\tr \comm{\comm{\comm{\W}{\Z}}{\Zb}}{\Wb} \comm{\Wb}{\comm{\Zb}{\comm{\Z}{\W}}}
                                                  +  \tr \comm{\comm{\comm{\W}{\Z}}{\Wb}}{\Zb} \comm{\Zb}{\comm{\Wb}{\comm{\Z}{\W}}} \\
                                                  - &\tr \comm{\comm{\comm{\W}{\Z}}{\Zb}}{\Zb} \comm{\Wb}{\comm{\Wb}{\comm{\Z}{\W}}}
                                                  -  \tr \comm{\comm{\comm{\W}{\Z}}{\Zb}}{\Wb} \comm{\Zb}{\comm{\Wb}{\comm{\Z}{\W}}} \bigr) : \; .
\end{split}
\ee
\end{subequations}


\section{Planar limit}

We now take the planar limit of the above $\grSU(2)$ dilatation operator $D$, i.e. we derive a new operator (again called $D$), which yields the dilatation matrix in 
the limit $N\to\infty$ with $\Lambda = \mbox{fixed}$. Since in the planar limit single and multi-trace states do not mix, we may and we do restrict the action of $D$ 
to the closed subset of single trace states.

Hence, the dilatation operator in the planar limit may be written as a sum of permutations of the fields in the trace. We use the notation of \cite{BKS}: Let 
$P_{k_1,k_2}$ exchange the fields at sites $k_1$ and $k_2$ in the trace (where the sites are periodically identified) and define
\be
  \{n_1,n_2,\ldots\} := \sum_{k=1}^L P_{k+n_1,k+n_1+1} P_{k+n_2,k+n_2+1} \cdots \; ,
\ee
where $L$ is the number of fields the operator is applied to, i.e. the length of the trace respectively chain. 
The identity operator is denoted by~$\{\}$. There are many trivial relations such as
\begin{align}
  \{\ldots,n,n,\ldots\} & = \{\ldots,\ldots\}      \\
  \{\ldots,n,m,\ldots\} & = \{\ldots,m,n,\ldots\}  \qquad\qquad \mbox{for $\modulus{n-m} \ge 2$} \\
  \{n_1,n_2,\ldots\}    & = \{n_1+m,n_2+m,\ldots\}
\end{align}
as well as the identity
\be
\begin{split}
  & \{\ldots,\ldots\} + \{\ldots,n\pm1,n,\ldots\} + \{\ldots,n,n\pm1,\ldots\} \\
  & - \{\ldots,n,\ldots\} - \{\ldots,n\pm1,\ldots\} - \{\ldots,n,n\pm1,n,\ldots\} = 0
\end{split}
\ee
which only holds in the $\grSU(2)$ subsector and expresses the fact that two objects cannot be placed completely antisymmetric onto three sites.

In order to find the planar version of the dilatation operator, we have made an ansatz for all independent permutations and determined their coefficients such that 
the $N\to\infty$ limit of \eqref{eqn:dilatation_operator} is correctly captured. We find the rather compact result:
\begin{subequations} \label{eqn:planar_dilatation_operator_matrix_theory}
\begin{align}
  D_0 & = \{\} \; , \\
  D_2 & = 2\{\} - 2\{0\} \; , \\
  D_4 & = -15\{\} + 19\{0\} - 2(\{0,1\} + \{1,0\}) \; , \label{D4} \\ 
  D_6 & = 187\{\} - 259\{0\} + 38(\{0,1\} + \{1,0\}) \nonumber \\
      & \quad + 4\{0,2\} - 4(\{0,1,2\} + \{2,1,0\}) - 2(\{0,2,1\} - \{1,0,2\}) \; . 
\end{align}
\end{subequations}
The presence of the antisymmetric term $(\{0,2,1\} - \{1,0,2\})$ in the three-loop part might appear unexpected. However, it is just a consequence of the fact that 
the energy operator \eqref{eqn:energy_operator} is not Hermitian\footnote{As a matter of fact $T$ is quasi-Hermitian, i.e. there exists an operator $S$ such that 
$T^\dagger= S^{-1}\, T\, S$.}. By a change of basis one can remove this term: after the following similarity transformation
\be
  D'(\Lambda) := e^{-\frac{\Lambda}{2} \{0\}} D(\Lambda) e^{\frac{\Lambda}{2} \{0\}}
\ee
we obtain $D'_0 = D'_0$, $D'_2 = D_2$, $D'_4 = D_4$ and
\be
  D'_6 = 187\{\} - 259\{0\} + 38(\{0,1\} + \{1,0\}) + 4\{0,2\} - 4(\{0,1,2\} + \{2,1,0\}) \; .
\ee
In \cite{us} we observed that the two-loop dilatation operator \eqref{D4}
failed to agree with the two-loop dilatation operator of SYM if one relates $M$ to $\gym$
according to \eqref{eqn:relation_coupling_constants_oneloop}. However, if one allows a
renormalization of this relation, then agreement with the full (planar and non-planar)
two-loop dilatation operator of SYM can be achieved. 
Moreover, we discover that this is true for the planar three-loop operator as well. Let thus $G_r$ be the renormalized coupling related to $G$ by 
\be
  G^2 = G_r^2 + \frac{7N}{2} G_r^4 - 11N^2 G_r^6 + \mathcal{O}(G_r^8) \; .
\ee
Then the dilatation operator $D'$ written in terms of $\Lambda_r := G_r^2 N$ becomes
\begin{subequations}
\begin{align}
  D'_{r,0} & = \{\} \; , \\
  D'_{r,2} & = 2\{\} - 2\{0\} \; , \\
  D'_{r,4} & = -8\{\} + 12\{0\} - 2(\{0,1\} + \{1,0\}) \; , \\
  D'_{r,6} & = 60\{\} - 104\{0\} + 24(\{0,1\} + \{1,0\}) + 4\{0,2\} - 4(\{0,1,2\} + \{2,1,0\}) \; ,
\end{align}
\end{subequations}
which is exactly the proposed field theory result, cf. equation (F.3) of \cite{BKS}. There the planar three-loop dilatation operator of SYM was determined using 
integrability as an input. Here we find it without any assumption.


\section{Integrability}

Since the planar dilatation operator \eqref{eqn:planar_dilatation_operator_matrix_theory} of plane-wave matrix theory in the $\grSU(2)$ subsector can be transformed 
to the corresponding dilatation operator of SYM, we know from the investigations of \cite{BKS} that it is integrable. I.e. there exists a charge
\be
  U(\Lambda) = \sum_{k=1}^\infty \Lambda^k U_{2k}\label{Udef}
\ee
which commutes with the dilatation operator. This charge pairs up states of different parity and is 
therefore responsible for their degenerate energies. Here parity refers to the operation of inverting 
the order of fields in
a trace, e.g. $P \tr(Z^3\phi^2Z\phi) = \tr( \phi Z\phi^2Z^3)$. The dilatation operator commutes with parity, which
is the reason for states of opposite parity not mixing. A priori there is no reason for a relation between the
spectra of positive and negative parity states. In the matrix model, however, we discover that there is a maximal
degeneracy of these spectra which we have explicitly evaluated up to dimension 10: 
Each representation that admits both parities has maximal degeneracy, see figure 1. 

\begin{figure}[t] \begin{center}
\psfrag{Dimension}[tc][Bc]{\small Dimension $\Delta(\Lambda=0.125)$}
\psfrag{4}[c][c]{\footnotesize $4$}
\psfrag{5}[c][c]{\footnotesize $5$}
\psfrag{6}[c][c]{\footnotesize $6$}
\psfrag{7}[c][c]{\footnotesize $7$}
\psfrag{8}[c][c]{\footnotesize $8$}
\psfrag{9}[c][c]{\footnotesize $9$}
\psfrag{10}[c][c]{\footnotesize $10$}
\psfrag{202}[c][c]{\scriptsize $[2,0,2]$}
\psfrag{212}[c][c]{\scriptsize $[2,1,2]$}
\psfrag{222}[c][c]{\scriptsize $[2,2,2]$}
\psfrag{303}[c][c]{\scriptsize $[3,0,3]$}
\psfrag{232}[c][c]{\scriptsize $[2,3,2]$}
\psfrag{313}[c][c]{\scriptsize $[3,1,3]$}
\psfrag{242}[c][c]{\scriptsize $[2,4,2]$}
\psfrag{323}[c][c]{\scriptsize $[3,2,3]$}
\psfrag{404}[c][c]{\scriptsize $[4,0,4]$}
\psfrag{252}[c][c]{\scriptsize $[2,5,2]$}
\psfrag{333}[c][c]{\scriptsize $[3,3,3]$}
\psfrag{414}[c][c]{\scriptsize $[4,1,4]$}
\psfrag{262}[c][c]{\scriptsize $[2,6,2]$}
\psfrag{343}[c][c]{\scriptsize $[3,4,3]$}
\psfrag{424}[c][c]{\scriptsize $[4,2,4]$}
\psfrag{505}[c][c]{\scriptsize $[5,0,5]$}
\includegraphics*[width=\textwidth]{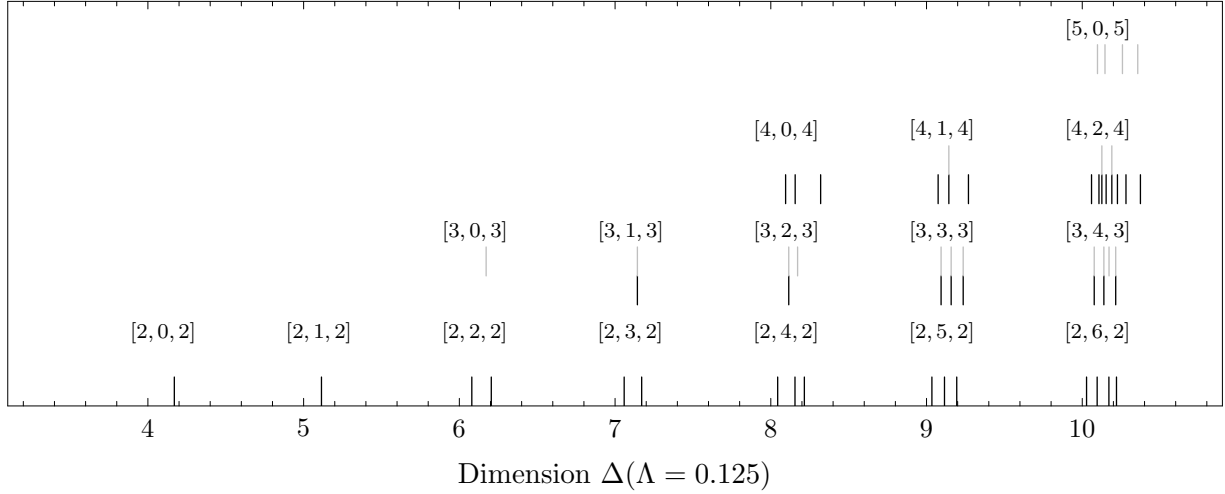} \end{center}
\caption{\small The complete spectrum of unprotected states in the $\grSU(2)$
subsector up to bare dimension 10 at three-loop order for $\Lambda=0.125$. 
States of positive (negative) parity 
are depicted as dark (light) lines. The states are grouped according to
the Dynkin labels of their representation. Note that whenever
states of positive and negative parity exist
within a representation, then they are degenerate. The explicit
results for the energy shifts are spelled out in the apppendix.} \label{tab:spectrum} \end{figure}

It is instructive to see this explicitly in the two lowest dimensional
representations with $\grSO(6)$ Dynkin labels $[3,1,3]$ and $[3,2,3]$. In the 
$[3,1,3]$ representation of $\Delta_0=7$ one has two single-trace operators of opposite
parity
\begin{align*}
{\cal O}_{+}^{[3,1,3]} & =       2 \tr[\W\W\W\Z\Z\Z\Z] - 3 \tr[\W\W\Z\W\Z\Z\Z] + 2 \tr[\W\W\Z\Z\W\Z\Z] \\
                 & \quad - 3 \tr[\W\W\Z\Z\Z\W\Z] + 2 \tr[\W\Z\W\Z\W\Z\Z] \; ,\\
{\cal O}_{-}^{[3,1,3]} & =     -   \tr[\W\W\Z\W\Z\Z\Z] +   \tr[\W\W\Z\Z\Z\W\Z] \; .
\end{align*}
The associated energy shifts up to order $\Lambda^6$ read
\be
\delta\Delta_{+} = \delta\Delta_{-} = 10 \Lambda^2 - 65 \Lambda^4 + 765 \Lambda^6
\ee
and are degenerate.

Similarly for the representation $[3,2,3]$ of $\Delta_0=8$ one finds one
single-trace operator of positive parity
\be
{\cal O}_{+}^{[3,2,3]}  =  \tr[\W\W\W\Z\Z\Z\Z\Z] - \tr[\W\W\Z\W\Z\Z\Z\Z] - \tr[\W\W\Z\Z\Z\Z\W\Z] + \tr[\W\Z\W\Z\Z\W\Z\Z] \nonumber \\
\ee
and two operators of negative parity
\begin{align*}
{\cal O}_{-,1}^{[3,2,3]} & = -\tr[\W\W\Z\W\Z\Z\Z\Z] + \tr[\W\W\Z\Z\W\Z\Z\Z] - \tr[\W\W\Z\Z\Z\W\Z\Z] + \tr[\W\W\Z\Z\Z\Z\W\Z] \; , \\
{\cal O}_{-,2}^{[3,2,3]} & = -\tr[\W\W\Z\W\Z\Z\Z\Z] - \tr[\W\W\Z\Z\W\Z\Z\Z] + \tr[\W\W\Z\Z\Z\W\Z\Z] + \tr[\W\W\Z\Z\Z\Z\W\Z] \; .
\end{align*}
In order to compute the energy shifts in the negative parity sector one has to diagonalize a $2\times 2$
matrix, ${\cal O}_{-}:=({\cal O}_{-,1}^{[3,2,3]},{\cal O}_{-,2}^{[3,2,3]})^\top$
\begin{align}
D_2 {\cal O}_{-} & = \left( \begin{matrix}  12 &  0 \\  0 &   8 \end{matrix} \right) {\cal O}_{-} &
D_4 {\cal O}_{-} & = \left( \begin{matrix} -78 & -4 \\ -4 & -48 \end{matrix} \right) {\cal O}_{-} &
D_6 {\cal O}_{-} & = \left( \begin{matrix} 938 & 60 \\ 68 & 540 \end{matrix} \right) {\cal O}_{-} \, .
\end{align}
The outcome of this straightforward exercise are the energy shifts
\begin{align}
\delta\Delta_{+} & = 8 \Lambda^2 - 48 \Lambda^4 + 536 \Lambda^6 \\
\delta\Delta_{-} & = \left\{\begin{matrix}
                     8 \Lambda^2 - 48 \Lambda^4 + 536 \Lambda^6 \\
                     12 \Lambda^2 - 78 \Lambda^4 + 942 \Lambda^6 
                     \end{matrix}\right.    
\end{align}
which again display degeneracy of opposite parity states.

Responsible for this degeneracy is the charge $U$ of \eqref{Udef}, which commutes with $H$. Up to three-loop orders
it is given by
\begin{subequations}
\begin{align}
  U_2 & = 4(\{1,0\} - \{0,1\}) \; ,   \\
  U_4 & = 8(\{2,1,0\} - \{0,1,2\}) \; , \\
  U_6 & = 8(\{1,0,3\} - \{0,1,3\} + \{0,3,2\} - \{0,2,3\}) + 18(\{3,2,1,0\}-\{0,1,2,3\}) \nonumber \\
      & \quad + 38\{0,1,3,2\} - 30\{2,1,0,3\} - 38\{0,3,2,1\} + 30\{1,0,2,3\}\nonumber\\
&\quad  - 34(\{1,0,3,2\} - \{0,2,1,3\}) \; .
\end{align}
\end{subequations}


\section{Nonsupersymmetric Deformations}

In this final section we turn to the question of how strongly the observed three-loop
integrability depends on the details of the model. A number of deformations are easily
implemented by switching of fields in our computer programs.

To begin with let us consider the purely bosonic $\grSO(3)\times \grSO(6)$ model by 
setting the fermions to zero in \eqref{hamiltonian}.
We calculate the effective vertex for this bosonic model similar to the above computation but subtract off
the vacuum energy shift -- which due to lack of supersymmetry does not vanish -- by 
hand\footnote{The vacuum energy shift is actually dominant for large $N$.}. 
Then the planar dilatation generator reads:
\begin{subequations}
\begin{align}
  D_0 & = \{\} \; , \\
  D_2 & = 14\{\} - 2\{0\} \; , \\
  D_4 & = -558\{\} + 122\{0\} - 4(\{0,1\} + \{1,0\}) \; , \\ 
  D_6 & = \tfrac{246929}{6} \{\} - \tfrac{63683}{6} \{0\} + 584 (\{0,1\} + \{1,0\})\nonumber\\
& \quad  +16 \{0,2\} - 16 (\{0,1,2\} + \{2,1,0\}) - 8 (\{0,2,1\} - \{1,0,2\}) \; . 
\end{align}
\end{subequations}
It is important to note that the planar parity pairs are degenerate under the 
{\it individual} action of the operators 
\begin{align*}
& \{\}, \quad  \{0\}, \quad \{0,1\} + \{1,0\} \quad \mbox{and} \\
& 2\{0,2\} - 2(\{0,1,2\} + \{2,1,0\}) - (\{0,2,1\} - \{1,0,2\}) \, .
\end{align*}
Hence the purely bosonic $\grSO(3)\times \grSO(6)$ model displays integrability up to
three-loops as well!

It turns out to be rather hard to break the hidden integrability structure of a
large $N$ matrix quantum mechanical model. For this we have studied a ``brutal''
deformation and considered the bosonic $\grSO(d)$ model ($i=1,\ldots,d$)
\be
  H = \frac{M}{2} \tr a_i^\dag a_i - \frac{\alpha}{2} \tr X_i X_j X_i X_j + \frac{\beta}{2} \tr X_i X_i X_j X_j
\label{brutal}
\ee
which for parameters $\alpha\neq\beta$ has lost all its traces of a higher dimensional
gauge field theory origin. Then one finds (again dropping the vacuum energy shift)
\begin{subequations}
\begin{align}
  D_0 & = \{\} \; , \\
  D_2 & = \left(-4\alpha + (2d+3)\beta \right) \{\} - 2\alpha \{0\} \; , \\
  D_4 & = \tfrac{1}{2} \left( (-100-12d)\alpha^2 + (168+96d) \alpha\beta + (-59-74d-24d^2) \beta^2 \right) \{\} \nonumber \\
      & \quad - 2 \left( 16\alpha^2 -(17+8d) \alpha\beta \right) \{0\} - 4 \alpha^2 (\{0, 1\} + \{1, 0\}) \; ,
\end{align}
\begin{align}
  D_6 & = \tfrac{1}{2} \bigl( (-560d-2440)\alpha^3 + (280d^2+3668d+6044)\alpha^2\beta \nonumber \\
      & \quad \qquad + (-1536d^2-5274d-4592)\alpha\beta^2 + (256d^3+1199d^2+1910d+1029)\beta^3 \bigr) \{\} \nonumber \\
      & \quad + \left( (-36d-832)\alpha^3 + (768d+1700)\alpha^2\beta + (-192d^2-797d-791)\alpha\beta^2 \right) \{0\} \nonumber \\
      & \quad + \left( -164\alpha^3 + 4(20d+37)\alpha^2\beta \right) (\{0, 1\} + \{1, 0\}) \nonumber \\
      & \quad + 16\alpha^2\beta \{0, 2\} - 16\alpha^3 (\{0, 1, 2\} + \{2, 1, 0\}) - 8\alpha^3 (\{0, 2, 1\} - \{1, 0, 2\}) \; . \label{here}
\end{align}
\end{subequations}
We see that the integrability of the model \eqref{brutal} only breaks down at the three-loop
order, as the last line of \eqref{here} destroys integrability for $\alpha\neq\beta$. If one
chooses $\alpha=\beta$, however, integrability is stable up to three-loops.

We interpret the findings of this section as an indication that  integrability  
might be a generic effect in large $N$ {\it gauge} quantum mechanical models 
and independent of supersymmetry. In the same instance we learn from the example \eqref{brutal}, 
that if integrability is to break down in a concrete model, it does so at rather high loop-orders (here three). 
This should caution us in our expectancy of an exact integrability of plane-wave matrix theory
or $\Ncal=4$ Super Yang-Mills based on three-loop results, even if the idea seems too appealing to
be false.

\bigskip
\leftline{\bf Acknowledgments}

\noindent We would like to thank Gleb Arutyunov,
Nakwoo Kim, Ari Pankiewicz, Matthias Staudacher and in particular
Niklas Beisert for very useful discussions.


\newpage
\section*{Appendix}

\begin{table}[h]
\begin{center}
\begin{tabular}{|c|c|l|l|}
\hline
$\Delta_0$ & rep       & $\delta\Delta_+$                            & $\delta\Delta_-$ \\ \hline
$4$        & $[2,0,2]$ & $12     \Lambda^2 - 90     \Lambda^4 + 1098   \Lambda^6$   & \\ \hline
$5$        & $[2,1,2]$ & $ 8     \Lambda^2 - 52     \Lambda^4 + 588    \Lambda^6$   & \\ \hline
$6$        & $[2,2,2]$ & $5.5279 \Lambda^2 - 30.987 \Lambda^4 + 326.57 \Lambda^6$   & \\ 
           &           & $14.472 \Lambda^2 - 107.01 \Lambda^4 + 1327.4 \Lambda^6$   & \\
           & $[3,0,3]$ &                                                            & $12 \Lambda^2 - 78 \Lambda^4 + 930 \Lambda^6$ \\ \hline
$7$        & $[2,3,2]$ & $4      \Lambda^2 - 20     \Lambda^4 + 202.5  \Lambda^6$   & \\
           &           & $12     \Lambda^2 - 84     \Lambda^4 + 997.5  \Lambda^6$   & \\
           & $[3,1,3]$ & $10     \Lambda^2 - 65     \Lambda^4 + 765    \Lambda^6$   & $10 \Lambda^2 - 65 \Lambda^4 + 765 \Lambda^6$ \\ \hline
$8$        & $[2,4,2]$ & $3.0121 \Lambda^2 - 13.863 \Lambda^4 + 137.83 \Lambda^6$   & \\
           &           & $15.208 \Lambda^2 - 112.68 \Lambda^4 + 1412.8 \Lambda^6$   & \\
           &           & $9.7802 \Lambda^2 - 63.455 \Lambda^4 + 719.41 \Lambda^6$   & \\
           & $[3,2,3]$ & $8      \Lambda^2 - 48     \Lambda^4 + 536    \Lambda^6$   & $8 \Lambda^2  - 48 \Lambda^4 + 536 \Lambda^6$ \\
           &           &                                                            & $12 \Lambda^2 - 78 \Lambda^4 + 942 \Lambda^6$ \\
           & $[4,0,4]$ & $6.4912 \Lambda^2 - 30.275 \Lambda^4 + 293.55 \Lambda^6$   & \\
           &           & $22.604 \Lambda^2 - 167.80 \Lambda^4 + 2059.3 \Lambda^6$   & \\
           &           & $10.904 \Lambda^2 - 69.927 \Lambda^4 + 859.19 \Lambda^6$   & \\ \hline
$9$        & $[2,5,2]$ & $2.3432 \Lambda^2 - 10.159 \Lambda^4 + 100.35 \Lambda^6$   & \\
           &           & $8      \Lambda^2 - 48     \Lambda^4 + 552    \Lambda^6$   & \\
           &           & $13.657 \Lambda^2 - 97.841 \Lambda^4 + 1189.7 \Lambda^6$   & \\ 
           & $[3,3,3]$ & $6.4532 \Lambda^2 - 35.770 \Lambda^4 + 382.11 \Lambda^6$   & $6.4532 \Lambda^2 - 35.770 \Lambda^4 + 382.11 \Lambda^6$ \\
           &           & $11.040 \Lambda^2 - 72.143 \Lambda^4 + 858.37 \Lambda^6$   & $11.040 \Lambda^2 - 72.143 \Lambda^4 + 858.37 \Lambda^6$ \\
           &           & $16.507 \Lambda^2 - 113.09 \Lambda^4 + 1356.5 \Lambda^6$   & $16.507 \Lambda^2 - 113.09 \Lambda^4 + 1356.5 \Lambda^6$ \\
           & $[4,1,4]$ & $ 5.072 \Lambda^2 - 22.574 \Lambda^4 + 220.97 \Lambda^6$   & \\
           &           & $10     \Lambda^2 - 65     \Lambda^4 + 785    \Lambda^6$   & $10     \Lambda^2 - 65     \Lambda^4 + 785    \Lambda^6$ \\ 
           &           & $18.928 \Lambda^2 - 133.43 \Lambda^4 + 1591.0 \Lambda^6$   & \\ \hline
\end{tabular}
\end{center}
\caption{Energy shifts for all unprotected states in $\grSU(2)$ subsector up to dimension $\Delta_0=9$.}
\end{table}


\end{document}